\documentstyle[12pt]{article}
\begin {document}
% \newcommand {\beq}{\begin{equation}}
% \newcommand {\eeq}{\end{equation}}
% \newcommand {\Lm}{\Lambda}
% \newcommand {\e}{\epsilon}
% \newcommand {\bt}{\bigtriangleup}
% \newcommand {\bd}{\Diamond}
% \newcommand {\cl}{\clubsuit}
% \newcommand {\ov}{\overbrace}
% \newcommand {\un}{\underbrace}
% \newcommand {\nn}{\normalsize}
% \newcommand {\lll}{\large}
% \newcommand {\LL}{\Large}
% \newcommand {\ft}{\footnotesize}
% \newcommand {\scr}{\scriptsize}
%%%%%%%%%%%%%%%%%%%%%%%%%%%%%%%%%%%%%%%
\title{Dynamical Scaling from Multi-Scale Measurements}
\author{
N. Persky\thanks{ Emails: nathanp@vms.huji.ac.il, benav@alon.cc.biu.ac.il ,
sorin@vms.huji.ac.il.} \\
Racah Institute of Physics,
Hebrew University, Jerusalem 91904, Israel \\
R. Ben-Av \\
Department of Physics, Bar-Ilan University, Ramat-Gan 52100, Israel \\
S. Solomon\thanks{ permanent address: Racah Institute of Physics,
Hebrew University, Jerusalem 91904, Israel } \\
Scuola Internationale Superiore di Studi Avanzati
SISSA, 34013 Trieste, Italy}
\date{Submitted to PR, 9 Jun, 1994. Report-no: RI-10  Jun, 1994}
\maketitle
\begin{abstract}
We present a new measure of the Dynamical
Critical behavior: the "Multi-scale Dynamical Exponent (MDE)",
$z^{md}$.
Using Dynamical RG concepts we
study the relaxation times, $\tau_{\Lambda} $, of a family of space scales
defined by $\Lambda = {1\over k}$.
Assuming dynamical universality we argue, and verify
numerically,
that $z^{md}$ has
the same value as the usually defined $z$.
We measure $z^{md}$
in the
2D Ising model
with the Metropolis and cluster dynamics
and find $z_{\rm{met}}^{md}=2.1\pm0.1$
and $z_{\rm{wolff}}^{md}=0\pm0.15$, respectively.
We note that in our approach $z^{md}$ is measured
using a single temperature and a single lattice size.
In addition,
in the Metropolis case we present a new method which helps
to overcome critical slowing down in the dynamical
measurements themselves.
\end{abstract}
%%%%%%%%%%%%%%%%%%%%%%%%%%%%%%%%%%%%%%%%%%%%%%%%%%

The static critical phenomena are well understood in simple
models and their properties can be calculated.
However, the knowledge of the time-dependence properties \cite{h}
is not in the same stage.
The various dynamical exponents for relaxation times,
 have not been related to the static exponents or calculated exactly
 even in the 2D Ising model.
Several kinds of analytical treatments (mainly for local classes of
dynamics) such as: dynamic Monte Carlo RG \cite{1,stauf},
 high temperature series expansion \cite{2} and $\epsilon$ expansion \cite{h},
have been employed
 in order to resolve this problem.
Others tried to estimate z using general theoretical approaches,
such as the recent Alexandrowicz conjecture \cite{4},
which expresses z as a combination
of the usual static critical exponents and a new geometric exponent.
Taking the 2D Ising model with one of its local dynamics, as a simple
example, one finds a wide range of theoretical and numerical
estimates for z (for a review and a list of
references see \cite{1}). The values quoted
there range from 1.82 to 2.24; so, although it seem that the
value around $2.05-2.2$ has a wide agreement, we can safely state that
there is no consensus in this issue
(the latest result \cite{tatata} is $2.163 (6)$
\footnote{We thank D. Stauffer for bringing it to our attention.}).

	From the practical point of view, z has a significant
 importance as well.
In order to get a sample of independent measurements one
must wait until the the system decorrelate from measurement
to measurement. In many situations, especially
near criticality, this takes a long time.
This Critical Slowing Down (CSD) \cite{h} phenomenon is one of the important
subjects one has to deal with while developing Monte Carlo (MC) algorithms.
In the last decade a family of new "global" dynamics that overcome
(at least partially) the CSD, was developed \cite{SW,D,W,R}.
The basic idea of the global methods is the ability to work on
'the relevant degrees of freedom'. This concept, is now used in a wide range
 of applications (for recent surveys see \cite{5}). Numerically
it was shown that those dynamics are not in the
universality class of the local dynamics, (such as Metropolis or
heat-bath), and their $z$'s may differ.
Since the key to improve the efficiency of the dynamics is reducing the
autocorrelation, estimation of the proper
$z$
is a crucial tool.
However, the value of
the dynamical critical exponent in those dynamics is not well
understood, and in fact only the lower bound has been
obtained analytically in some cases \cite{socal}.

	Measuring z is a difficult task, a review of common problems and difficulties
can be found in refs. \cite{1,6}.
Although the usual definition of $z$ is : $ \tau \propto \xi^z$,
in practice, due to technical problems, other ways of calculating z
are usually preferred.
Many standard methods for measuring z, involve repeating the measurements
of $\tau$ for various lattices with different linear sizes, L. Then, using
finite-size arguments one deduces $z$ from the relation $ \tau \propto L^z$
\cite{7}.
One of the problems with those methods is the dependence
of the critical temperature on $L$.

	The main idea of this letter, is that a system at criticality
contains all the possible time and length scales.
Studying the internal relations
between the spatial and the time scales in one system, would
yield the desired knowledge on the dynamical properties.
following \cite{bb},
we define the elements of the autocorrelation matrix to be:
\begin{equation}
M_{\bf{l}, \bf{j}}(t)=
<S_{\bf{l}}(0)S_{\bf{j}}(t)>-<S_{\bf{l}}><S_{\bf{j}}>.
\end{equation}
where $\bf{l}$ and $\bf{j}$ are points on the lattice
%\newline
(e.g in the 2-D lattice
${\bf l}=(n_x,n_y)~ ; 1\leq n_x \leq L_x, 1\leq n_y \leq L_y$ ),
and $S_{\bf{l}}$
is any local observable (e.g. the spin at the point $\bf{l}$).
In the case of a translation invariant dynamics one has:
\begin{equation}
M_{\bf{l}+\bf{d},\bf{j}+\bf{d}}(t)=M_{\bf{l},\bf{j}}(t)
\end{equation}
where $\bf{d}$ is any lattice vector. Thus, one can identify the eigenvectors
of the
autocorrelation matrix as Fourier modes. Each
eigenvalue $\lambda_{\bf{k}}(t) $ corresponding to the
eigenvector $ e^{i \bf{j} \cdot \bf{k}}$ satisfies:
\begin{equation}
\sum_{\bf{j}} M_{\bf{l},\bf{j}}(t) e^{i \bf{j} \cdot \bf{k}}=
\lambda_{\bf{k}}(t) e^{i \bf{l} \cdot \bf{k}}.
\end{equation}
 For each wave-vector $\bf{k}$ one can assign a decorrelation time
$\tau_{\bf{k}}$ via:
\begin{equation}
\lambda_{\bf{k}}(t)\sim e^{-t/\tau_{\bf{k}}}, ~~~~~~
 t\rightarrow\infty
\end{equation}
Note that for instance,
 in the spin case,
$\tau_{\bf 0}$ ($\tau$ for ${\bf k}=(0,0))$
is identical to the usually defined
exponential autocorrelation time for the magnetization.

Measuring the decay of the eigenvalue, $\lambda_{\bf{k}}(t) $,
we estimate $z^{md}$ assuming the dynamical RG \cite{ma} scaling relation:
\begin{equation}
 \tau_{k}\propto k^{-z^{md}}.
\end{equation}
where $k$ denotes the norm of $\bf{k}$.
The reason for this assumption is that the corresponding length scale
of a Fourier mode is:
$\Lambda\propto k^{-1} $. Therefore assuming dynamical universality
one expects that $ \tau_k\propto \Lambda^{z^{md}}$.

\em
The LAG Algorithm \rm {\Large\bf$-$}The
measurement of a \em dynamical observable \rm associated with
 an algorithm with local dynamics, just as any
measurement of a static property, suffers from CSD. Namely, even in the
measurements of the CSD itself,
one has to wait a long time,
in order to avoid correlations between successive
measurements of the autocorrelation matrix {\bf M}(t).
To overcome this difficulty we introduce the
"Local After Global" (LAG) technique. The LAG combines a global dynamics
(GD) (which does not suffer from
CSD) with a local dynamics (LD) such that the LD is used
only when it is needed for measurements. More precisely,
first we relax the system
with GD then
we preform LD steps as much as needed to have one measurement of
the dynamical observable (in our case {\bf M}(t)), then we preform again
 a few steps of GD in order to decorrelate
the system, then again the LD steps and so on.
In figure 1, we illustrate this procedure.

\em
	 The Simulation.\rm
{\Large\bf$-$}The model in which we preformed our simulation is
the 2D Ising on a square lattice \cite{IS}.
We measured the autocorrelation eigenvalues for both a GD (cluster) \cite{W}
and a LD (Parallel Checkboard Metropolis (PCM)\cite{note}) algorithm.
The lattice linear
sizes were L=64,128 \cite{note2},
and the temperature was
$T=T_c={2\over{log(\sqrt2+1)}}$.
In the PCM case we employed the cluster dynamics
to relax and to decorrelate the system according to the rules of the LAG
procedure.
In both cases we had periodic boundary conditions.
In the cluster dynamics since
the sites are randomly accessed, the translation
invariance of the dynamics is guaranteed.
In order to have translation
invariance in the PCM as well,
we define a block-spin
variables which is the sum of the spins on the
four sites of a 2$\times$2 square sub-lattice.
In both cases we discard the first 10,000 cluster sweeps and then
took $3\times10^5-2\times10^6$ different measurements of the
autocorrelation matrix, with ten cluster
sweeps between each measurement.
The translation invariance allows us to
limit the actual computation to only one line of the autocorrelation
matrix in each dimension. The reflection symmetry further reduces
 the actual necessary measurements to almost
half of a line in each dimension. Altogether
one has to measure only $ (L/4+1)^2$ elements. The other parts of
the matrix are actually, reflections and cyclic permutations of
 this 'quarter line'.
In practice at every step of a $M_{\bf{l}, \bf{j}}(t)$ measurement, we made a
spatial average by measuring
the matrix element for every one of 4 (16 for L=128) spin-blocks, and then
average over all
the measurements.
We repeated this process for 10 independent systems and estimated the
error bars by the standard deviation.
In order to estimate $\tau_{\bf k}$ we fit
the exponential in eq. (4), and in order to
deduce $z^{md}$ we fit to the  power law eq. (5).
The fits were done using nonlinear least square \cite{9}
which gives the error bars as well.

In figures 2 and 3 one can see a log-log plot of $\tau_{\bf{k}}(k)$, and the
evaluated
$z$ for the Metropolis and cluster dynamics, respectively.
In order to avoid transients in the calculation of $\tau_{\bf{k}}$ we
waited until the momentary $\tau_{\bf{k}}$ had been reasonably stabilized.
The final results of our fitting are
$z_{\rm{met}}^{md}=2.1\pm0.05$, $z_{\rm{wolff}}^{md}=0.1\pm0.2$,
for L=64,
and
$z_{\rm{met}}^{md}=2.1\pm0.1$, $z_{\rm{wolff}}^{md}=
0\pm0.15$,
for L=128.

Note the shift in emphasis  compared with the usual technique for measuring
dynamical critical exponents.
While usually, the main measurement is on the total magnetization
and its time evolution ($ \tau_{{\bf k}=(0,0)} \sim L^z$),
in our case we discard the low values of ${\bf k}$ as finite-size artifacts.
Incidentally these values are also the
most demanding in terms of running times.

Instead,
the  explicit \em simultaneous \rm measurement
of \em all \rm the scales below the maximal scale in the system
allows one to obtain the $z$ value within a highly self-consistent
measurement. The quite precise linearity of the
the graphs 2 and 3 verifies the consistency of the
$z$ values prevailing  at the various scales within the system.

As $L \rightarrow \infty$
the range of dynamically active scales within the system extends to infinity
and in fact this produces a power law.

In this letter we introduced
a new method to investigate the internal relations
between the spatial and the time scales in statistical models, and
showed that the the dynamical
RG ideas can be implemented practically.
The results in the 2D Ising model confirmed that the Dynamical RG assumption
for
the scaling relation
$ \tau_{\Lambda} \propto \Lambda^{z^{md}}$
holds both for the Metropolis and the cluster
dynamics,
suggesting that it does not depend on the particular kind of
dynamics.
Let us emphasize that our method
does not require runs at several $\xi$'s (corresponding to
different temperatures)
or at several lattice sizes $L$: we  perform
all the measurements relevant for all scales at fixed temperature
and fixed lattice size.

One should note that each Fourier subspace contains a set of many
$\tau$'s. That is the $\tau_{\bf{k}}(t)$ can be Laplace or
Fourier transformed \cite{fu} or directly be fitted \cite{2exp},
to be a linear combination of many decay-times.
However in our case we took the lowest mode in each
$k$ subspace and these lowest modes produced a power law
consistent with the dynamical universality assumption.
One can go farther and use our method to analyze the
internal structure of the Fourier subspaces as well
as seeing the correspondence between them.

In addition, we presented an efficient tool to analyze dynamical
properties without suffering from CSD. The LAG technique
combines the dynamics which one wants to analyze, with a global
dynamics used to decorrelate the system.
Using GD to decorrelate the system is of great
importance even when the main target of the simulation
is the properties of the local dynamics.
The relevance of LAG technique goes beyond the objectives
of the present letter and we intend to exploit it in
the future, in a wider range of applications.

Finally, we want to report on an interesting short time scales property of
autocorrelation functions.
On long time scales the autocorrelation function is convex,
and it is customary to expect that this property holds in general.
A sufficient condition for convexity is an existence of
spectral representation of the dynamic process.
Indeed, in many classes of dynamics this is actually the case.
In fact, in those dynamics which satisfy detailed balance,
the corresponding transition matrix is symmetric in the Hilbert space
$l^2(P_{eq})$ \cite{socal2}, namely the Hilbert space with the
measure (weight)
$P_{eq}$, the unique stationary distribution.
Therefore the spectral representation, and consequently
the convexity are guaranteed.
On contrary,
in dynamics which do not satisfy detailed balance
such as 'typewrite order' Metropolis, the convexity is not guaranteed.
In fact, in a few types of dynamics we were able to detect at
short times non convex
behavior of the autocorrelation function.
The convexity holds only after few steps.
We discuss this result elsewhere \cite{np}.

The research of NP and SS was
supported by
Germany Israel Foundation (GIF),
and by the
Foundation for Fundamental Research of the Israeli Academy.
The research of RBA
was supported by the ministry of Science and Technology,
Israel.

%%%%%%%%%%%%%%%%%%%%%%%%%%%

\begin{thebibliography}{99}
\bibitem{note} In fact in the Metropolis case we preformed simulation
in random chosen sites metropolis dynamics as well ,
and we got similar results.
\bibitem{note2}We use 2 lattice
sizes to show that the result is independent in L.
As explained before, the MDE procedure does not require in principle
results from
different L values.
\end{thebibliography}
\end {document}